\documentclass[preprint,12pt,authoryear]{elsarticle}

\usepackage{amssymb}
\usepackage{amsmath}
\newcommand\numberthis{\addtocounter{equation}{1}\tag{\theequation}}

\DeclareFontFamily{OT1}{pzc}{}
\DeclareFontShape{OT1}{pzc}{m}{it}{<-> s * [1.10] pzcmi7t}{}
\DeclareMathAlphabet{\mathpzc}{OT1}{pzc}{m}{it}

\newcommand{\tyf}{f}
\newcommand{\tyc}{\mathpzc{c}}

\journal{Astroparticle Physics}

\begin{document}

\defcitealias{REF::VASSILIEV::APP2007}{VFB}
\defcitealias{REF::BRETZ_RIBORDY::DC2013}{BR}
\defcitealias{REF::SCHLIESSER::THIRD_ORDER::APP2005}{SM}

\begin{frontmatter}



\title{The perfomance of generalized Davies-Cotton optical systems 
with infinitesimal mirror facets.}


\author{S.~Fegan}%[orcid=0000-0002-9978-2510]
\ead{sfegan@llr.in2p3.fr}
\address{Laboratoire Leprince-Ringuet, Ecole Polytechnique, CNRS/IN2P3, Palaiseau, France}

\begin{abstract}

This paper presents Taylor expansions for the imaging and timing characteristics of
spherical optical systems with infinitesimal mirror facets, sometimes referred
to as ``modified Davies-Cotton'' telescopes. Such a system comprises a
discontinuous spherical mirror surface whose curvature radius is different from
its focal length, and whose mirrors are aligned to suppress spherical
aberration. Configurations that range between two ``optima'' are, one of
which minimises tangential comatic aberration and the other that minimises
timing dispersion.

\end{abstract}


\end{frontmatter}


\section{Introduction}
\label{SEC::INTRODUCTION}

In Section~2 of \citet[hereafter
\citetalias{REF::VASSILIEV::APP2007}]{REF::VASSILIEV::APP2007} we presented an
expression for the first and second moments of the light distribution formed by
parallel rays reflected from an ``ideal'' Davies-Cotton telescope
\citep[DC]{REF::DAVIESCOTTON::1957}, i.e.\ one with infinitesimal facets,
neglecting diffraction. Our expression consisted of a Taylor expansion in
leading orders of the off-axis field angle, $\delta$, and in
$\frac{1}{4\tyf^2}$, where $f=F/D$ is the ratio of the focal distance, $F$, to
the diameter of the aperture, $D$. We discussed the leading order terms
expansion for the tangential and sagittal second moments, spherical 
aberration ($\propto\delta^0/f^6$), coma ($\propto\delta^2/f^4$) and astigmatism 
($\propto\delta^4/f^2$), and also considered the effects of focal-plane curvature. 
Additional terms in the leading-order expansion were subsequently given by 
\citet[hereafter \citetalias{REF::BRETZ_RIBORDY::DC2013}]{REF::BRETZ_RIBORDY::DC2013}.
\citet[hereafter \citetalias{REF::SCHLIESSER::THIRD_ORDER::APP2005}]
{REF::SCHLIESSER::THIRD_ORDER::APP2005} provided third-order expressions
for \textit{continuous} mirrors in the form of conics of rotation, which cover
parabolic, spherical, and elliptical mirrors. Such analytic approximations are
generally useful, for example to (1) quickly evaluate the performance of
tessellated optical systems, neglecting aberrations from the finite facet size,
and (2) validate DC ray-tracing simulation codes.

In VHE gamma-ray astronomy with imaging atmospheric \v Cerenkov telescopes
(IACTs) such simulation codes are often part of the Monte-Carlo simulation chain
from which the performance of the IACT instrument is calculated. For performance
reasons, such codes are not usually based on general-purpose ray-tracing
packages, rather they are custom-developed codes optimised for DC-like
telescopes. Expansions such as those in \citetalias{REF::VASSILIEV::APP2007} and
\citetalias{REF::BRETZ_RIBORDY::DC2013} could be used to validate such codes by
simulating telescopes with a large number of small (but finite) mirror facets.
However, these expressions apply to the case of a telescope with infinitesimal 
mirrors arranged in such a way that all incoming rays are reflected to the 
focus irrespective of field angle, whereas most DC telescopes that have been built, 
and most simulation codes, assume identical mirror facets arranged on 
a regular grid perpendicular to the optic axis, e.g. hexagonal or square. These 
facets cannot all be aligned perpendicular to the incoming rays 
and hence some rays will not be reflected, passing through ``the cracks'' 
between mirrors that arise due to this non-perpendularity, giving a small but 
significant difference between the analytic expressions and the limiting case
from the simulations.

Some recent IACTs, such as the medium-sized telescopes
\citep[MST]{REF::GARCZARCZYK::MST::ICRC2015} of the \v Cerenkov telescope array
(CTA), deviate from the classical DC design in which the curvature radius of the
spherical dish, $R$, equals the focal distance. The response of such
\textit{generalised DC} telescopes, and those that have a parabolic dish design,
cannot be approximated with the expressions of
\citetalias{REF::VASSILIEV::APP2007}. The expressions of
\citetalias{REF::SCHLIESSER::THIRD_ORDER::APP2005} apply to continuous mirror
systems intermediate between spherical and paraboloid, however these systems
have significantly worse optical performance than that of the discontinuous
generalised DC system, except in the parabolic case.

The following section presents expressions for the performance of a general 
spherical telescope with infinitesimal mirror facets laid out on a regular grid, 
which can be used for design studies and to validate ray-tracing simulations with 
small mirror facets. Since the methodology is very similar to that of 
\citetalias{REF::VASSILIEV::APP2007}, only differences between the analyses
are presented in detail.

\section{Taylor series expansions}
\label{SEC::CALCULATION}

Adopting the notation of Section~2 of
\citetalias{REF::VASSILIEV::APP2007}, the surface of
the primary mirror is $\vec
r(\varphi,\theta)=R(\sin\theta\cos\phi,\sin\theta\sin\phi,1-\cos\theta)$, with
$\theta\in[0,\sin^{-1}(D/2R)]$ and $\varphi\in[0,2\pi]$. For a general spherical
telescope there is no global alignment point; instead to focus parallel on-axis
rays to a point at $\vec{r_F}=(0,0,F)$ each facet must be aligned to a point on
the z-axis the same distance from the focal point as is the facet, $\vec {r_A} =
(0,0,z_A)$ with,
\begin{equation}\label{EQ::ALIGNMENT_PT}
  z_A=F+|\vec{r}(\phi,\theta)-\vec{r_F}|=F+\sqrt{4R(R-F)\sin^2(\theta/2)+F^2}.
\end{equation}
The normal for each facet is then $\vec n(\varphi,\theta) = \frac{\vec{r_A}-\vec
r(\varphi,\theta)}{|\vec{r_A}-\vec r(\varphi,\theta)|}$. As in
\citetalias{REF::VASSILIEV::APP2007} the direction of the 
incoming parallel rays is written as $\vec g=(\sin\delta,0,-\cos\delta)$,
and the direction of the reflected ray as $\vec u(\varphi,\theta)=\vec
g-2\left(\vec g\cdot\vec n(\varphi,\theta)\right)\vec n(\varphi,\theta)$. The
path of the reflected ray is from the mirror to the focal plane can be
parameterised by $s$ as, $\vec r_{ray}(s;\varphi,\theta) =
\left(x_{ray}(s),y_{ray}(s),z_{ray}(s)\right) = \vec r(\varphi,\theta)+s\vec
u(\varphi,\theta)$, from which the  propagation distance from the mirror to the
focal plane, $s_{fp}(\varphi,\theta)$, can be found by solving
$z_{ray}(s_{fp})=F$, where a curved focal plane is not considered. The 
coordinates of the ray on the focal plane are then 
$x_{fp}(\varphi,\theta)=x_{ray}(s_{fp}(\varphi,\theta); \varphi,\theta)$
and $y_{fp}(\varphi,\theta)=y_{ray}(s_{fp}(\varphi,\theta); \varphi,\theta)$.

Calculation the moments of the distribution of rays on the focal plane is
achieved by integrating the appropriate function over the infinitesimal 
mirrors of surface area
$\mathrm dS = R^2 \cos\theta\sin\theta\mathrm d\theta \mathrm d\phi$.
Each mirror facet presents an area of
$|\vec g\cdot\vec n(\varphi,\theta)|\mathrm dS$ to the incoming rays, which
are reflected to the focal plane, assuming no obscuration. The moments are
then calculated using the functional,
\begin{equation}\label{EQ::FUNCTIONAL}
\mathcal I[f(\varphi,\theta)] =
\int_{0}^{\sin^{-1}(D/2R)}\int_{0}^{2\pi} f(\varphi,\theta)
|\vec g\cdot\vec n(\varphi,\theta)| R^2 \cos\theta\sin\theta
\mathrm d\theta\mathrm d\phi,
\end{equation}
where the inclusion of the dot-product term represents the second difference
between this analysis and that of
\citetalias{REF::VASSILIEV::APP2007}.

The effective area of the telescope for rays is given by  $A=\mathcal I[1]$. The
first moments of the light distribution are then
$\langle x\rangle=\mathcal I[x_{fp}]/A$ and
$\langle y\rangle=\mathcal I[y_{fp}]/A$. The centered second order moments
are
$\langle \Delta x^2\rangle = \mathcal I[x_{fp}^2]/A-\langle x\rangle^2$,
$\langle \Delta y^2\rangle = \mathcal I[y_{fp}^2]/A-\langle y\rangle^2$, and
$\langle \Delta x\Delta y\rangle = \mathcal I[x_{fp}y_{fp}]/A-\langle x\rangle\langle y\rangle$.
Writing the propagation time of the rays through the system with respect to an
arbitrary offset as,
\[ t_{fp}(\varphi,\theta) = \frac{s_{fp}(\varphi,\theta) + \vec g\cdot\vec r(\varphi,\theta)}{\mathrm{c/n}}, \]
where $\mathrm{c/n}$ is the speed of light in the air, the variance of the distribution
of arrival times on the focal plane can be expressed as
$\langle \Delta t^2\rangle = \mathcal I[t_{fp}^2]/A -
\mathcal I[t_{fp}]^2/A^2$.

As in \citetalias{REF::VASSILIEV::APP2007} the integrals are computed and, with the 
exception of $A$, the results expressed as Taylor expansions in leading orders of 
$\delta$ and $\frac{1}{4\tyf^2}$. For convenience the results are expressed in 
terms of the ratio of the focal distance to reflector radius, $\tyc=F/R$. 
For the classic DC telescope, $\tyc=1$; a good approximation
to a parabolic reflector can be made with $\tyc=1/2$; the CTA MST corresponds to
$\tyc=16/19.2=0.83$ \citep[and $\tyf=16/12=1.33$,][]{REF::GARCZARCZYK::MST::ICRC2015}.

In the case of $A$ the dependency on $\delta$ can be expressed fully, so the
Taylor expansion is done only in terms of $\frac{1}{4\tyf^2}$,
\begin{equation}
  A = \pi\left(\frac{D}{2}\right)^2\left(1-\frac{1}{64\tyf^2}-\frac{16\tyc-11}{6144\tyf^4}\right)\cos\delta.
\end{equation}
This is smaller than the canonical value of $\pi \left(\frac{D}{2}\right)^2\cos\delta$ by
$1.6\%/\tyf^2$ to leading order, showing that with this mirror configuration
rays are lost even in the ``ideal'' case of an infinite number of identical 
infinitesimal facets on a grid.

The centroids of the light distribution on the focal plane are,
\begin{equation}\label{EQ::CENTROID_X}
\frac{\langle x\rangle}{F} =\,
\delta\left(1+\frac{3-2\tyc}{32\tyf^2}+\frac{-16\tyc^3+50\tyc-11}{6144\tyf^4}\right)
+ \frac{\delta^3}{3}\left(1+\frac{3-\tyc}{16\tyf^2}\right)
\end{equation}
and $\frac{\langle y\rangle}{F}=0$,
from which it is evident that there is a leading-order plate-scale correction of
$(3-2\tyc)/32\tyf^2$. This amounts to a correction of $3.1\%/\tyf^2$ for the
classic DC design, $6.3\%/\tyf^2$ for a parabolic mirror. In the case 
of $\tyc=3/2$, i.e.\ a system with $R=2F/3$, the first order plate-scale 
correction is eliminated.

The centered second order moments in the tangential ($x$) and sagittal ($y$)
directions are,
\begin{align*}
\frac{\langle\Delta x^2\rangle}{F^2} &=
\frac{\delta^2}{1024\tyf^4}\left(\frac{4\tyc^2-12\tyc+11}{3}
+ \frac{32\tyc^4 - 48\tyc^3 - 96\tyc^2 + 184\tyc - 13}{192\tyf^2}\right.\\
& \quad\quad\left. +\frac{2240\tyc^6 - 2592\tyc^5 - 5696\tyc^4 - 224\tyc^3 + 17026\tyc^2 - 4626\tyc + 1119}{92160\tyf^4} \right) \\
& + \frac{\delta^4}{16\tyf^2}\left(1 + \frac{8\tyc^2 - 36\tyc + 217}{576\tyf^2}
+ \frac{32\tyc^4 - 72\tyc^3 - 144\tyc^2 + 1276\tyc + 629}{18432\tyf^4}\right) \\
& + \frac{\delta^6}{12\tyf^2}\left(1 + \frac{17\tyc^2 - 111\tyc + 1528}{2880\tyf^2} \right),\numberthis\label{EQ::VAR_X}\\
\end{align*}
\begin{align*}
\frac{\langle\Delta y^2\rangle}{F^2} &=
\frac{\delta^2}{1024\tyf^4}\left(\frac{2}{3} + \frac{24\tyc + 11}{192\tyf^2}
+ \frac{1152\tyc^3 + 3456\tyc^2 + 4136\tyc - 811}{184320\tyf^4}\right) \\
& + \frac{\delta^4}{2304\tyf^4}\left(1 + \frac{24\tyc + 47}{128\tyf^2}
+ \frac{1152\tyc^3 + 3456\tyc^2 + 17960\tyc + 3185}{122880\tyf^4}\right) \\
& + \frac{17\delta^6}{69120\tyf^4}\left(1 + \frac{408\tyc + 1627}{2176\tyf^2} \right),\numberthis\label{EQ::VAR_Y}
%\frac{\langle\Delta x\Delta y\rangle}{F^2} =&\, 0\\
\end{align*}
while the covariance term is zero, $\frac{\langle\Delta x\Delta
y\rangle}{F^2}=0$. 
We note that spherical aberration ($\propto\delta^0/f^6$) is suppressed for 
all values of $\tyc$, by construction. The leading order aberration is coma 
($\propto\delta^2/f^4$). In the expression for the tangential coma, the
first sub-term (i.e. quadratic in $\tyc$) is twice as large in the parabolic 
reflector case ($2$ for $\tyc=1/2$) as it is in the DC case ($1$
for $\tyc=1$). This term is minimised by $\tyc=3/2$, the same value that was
found to eliminate the plate-scale correction above, for which the term is
$2/3$. The second sub-term of the tangential coma expression (the fourth order polynomial
$\propto\delta^2/\tyf^6$) is also close to its minimum at $\tyc=3/2$ (the
minimum is at $\tyc=1.48$). In fact, for $\tyc=3/2$ the tangential and sagittal
coma expressions are the same to two leading orders, both equalling
$\frac{\delta^2}{1024\tyf^4}\left(\frac{2}{3}+\frac{47}{192\tyf^2}\right)$.
However, the tangential astigmatism term is significantly larger than the
sagittal, which means that $\frac{\langle\Delta
x^2\rangle}{F^2}\gg\frac{\langle\Delta y^2\rangle}{F^2}$ for $\delta\gg0$, even
for $\tyc=3/2$.

The variance in the distribution of ray impact times on the focal plane is,
\begin{align*}
\langle\Delta t^2\rangle &= \left(\frac{D}{\mathrm c/\mathrm n}\right)^2
\left\{ \frac{1}{768\tyf^2}\left((2\tyc-1)^2 + \frac{(2\tyc-1)(2\tyc^3-2\tyc+1)}{8\tyf^2} \right.\right.\\
& \quad \left. + \frac{4480\tyc^6 - 1728\tyc^5 - 3840\tyc^4 - 1664\tyc^3 + 8196\tyc^2 - 5508\tyc + 1153}{61440\tyf^4}\right) \\
& - \left. \frac{\delta^2}{256\tyf^2}\left( 2\tyc-1 + \frac{3\tyc^3 + 6\tyc^2 - 5\tyc - 2}{24\tyf^2} \right)  \right\},\numberthis\label{EQ::VAR_T}
\end{align*}
which is, in general, non-zero even for $\delta=0$. A parabolic mirror 
is well known to be isochronous for rays parallel to the optic axis; for the 
spherical approximation, with $\tyc=1/2$, the first two terms of $\delta=0$ are 
identically zero, but the third term, of order $\frac{1}{\tyf^6}$, is marginally 
non-zero. To get an idea of the relative size of the $\delta^0$ and $\delta^2$ 
terms, keeping only the leading orders in $\frac{1}{4\tyf^2}$, the RMS of the 
impact time distribution for $\tyc>1/2$ can be approximated as,
\begin{equation} \label{EQ::RMS_T}
\sigma_t = \sqrt{\langle\Delta t^2\rangle} \approx
\frac{D}{10\,\mathrm{m}}\,\frac{\mathrm{n}}{\tyf} \left\{
  1.2036\,(2\tyc-1) - 0.00055\,\left(\frac{\delta}{1\,\mathrm{deg}}\right)^2
  \right\}\,\mathrm{ns},
\end{equation}
from which it is evident that the dependency on field angle is small
compared to the constant term, except for the configurations close to parabolic,
$\tyc\approx1/2$, where higher order terms dominate. The classic DC design ($\tyc=1$)
has a characteristic dispersion of $\sigma_t\approx1.2/\tyf$\,ns for a
$10$\,m telescope, whereas for the case of the ``minimal coma'' solution
with $\tyc=3/2$ the time dispersion is twice as large. For the case of
$\tyc=1/2$ the spherical approximation to a parabolic reflector has a
leading-order time dispersion of,
\[ \sigma_t\approx 0.019\frac{D}{10\,\mathrm{m}}\frac{\mathrm{n}}{\tyf^3}
\sqrt{1+0.38\tyf^2\left(\frac{\delta}{1\,\mathrm{deg}}\right)^2}
\,\mathrm{ns}. \]

Finally, the configuration with planar mirror surface, $\tyc=0$, has a 
time dispersion that is no worse that the DC, but the RMS of the tangential
coma is twice as large as the DC case. Large planar mirror systems are used
as solar collectors, but it seems unlikely that the configuration provides 
a compelling advantage for applications in large imaging telescopes. However, 
the solution is a close analogue of the Fresnel lens, which is a flat, 
discontinuous, refractive focussing element that has proven useful and 
cost-effective on smaller scales. A reflective equivalent would resemble a 
negative-curvature (diverging) Fresnel lens with a reflective coating.
The curvature on each of the grooves would be significantly less than for 
a lens of equal focal length, and the same diamond-turning or 3D-printing 
technology could be used to form the surface. In wideband imaging
applications, such as \v Cerenkov astronomy, such a reflector may provide
advantages over a Fresnel lens, whose imaging quality is limited by chromatic 
aberrations\footnote{Strictly the term $\vec g\cdot\vec n$ added in 
Section~\ref{SEC::CALCULATION} does not apply to a Fresnel-like reflector,
but the effect of including it is small.}.

\section{Accuracy}
\label{SEC::ACCURACY}

\begin{figure}[t]
\includegraphics[width=0.32\textwidth]{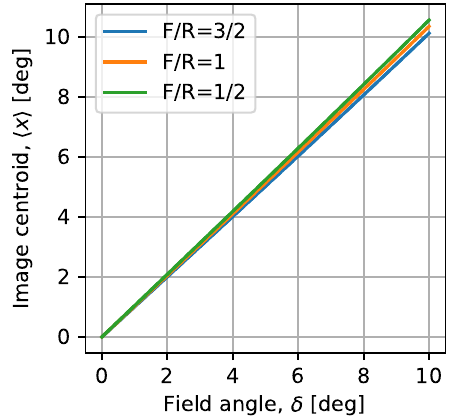}\hspace{\fill}%
\includegraphics[width=0.32\textwidth]{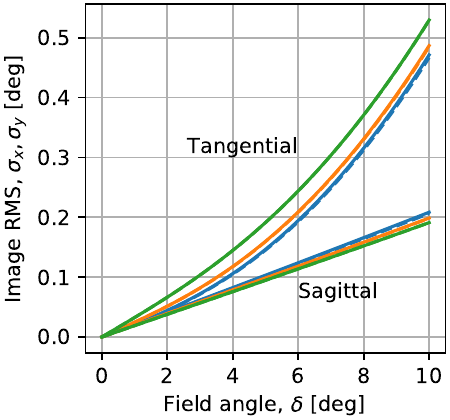}\hspace{\fill}%
\includegraphics[width=0.32\textwidth]{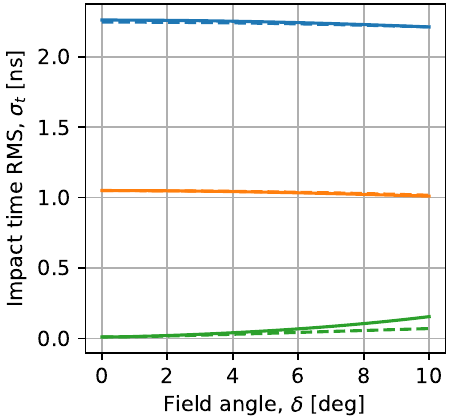}
\caption{\label{FIG::QUAD_TAYLOR_COMP} Comparison between Taylor series
expressions (dashed lines) and results from quadrature numerical integration
(solid lines) for: image centroid (left); tangential and sagittal image RMS
(middle); and arrival time dispersion (right).}
\end{figure}

Figure~\ref{FIG::QUAD_TAYLOR_COMP} shows a comparison between the Taylor
expansions given above and the results from numerically integrating
Equation~\ref{EQ::FUNCTIONAL} to high accuracy, for field angles of
$\delta<10^\circ$. The calculation assumes a system with $D=10$\,m, $F=12$\,m,
and three values for the curvature radius, $R=\{8,12,24\}$\,m, indicated as
$\tyc=F/R=\{3/2,1,1/2\}$ respectively on the plots. The three plots correspond to:
image centroid, Equation~\ref{EQ::CENTROID_X} (left); tangential and sagittal
image RMS, Equations~\ref{EQ::VAR_X} and \ref{EQ::VAR_Y} (middle); and arrival
time dispersion, Equation~\ref{EQ::VAR_T} (right).

The accuracy of the Taylor expansions depend on $\tyf$, $\tyc$, and $\delta$,
but for the cases presented here, all are accurate to better than $1\%$ with the
following two exceptions: (1) for $\tyc=3/2$ the error in $\sigma_x$ and
$\sigma_y$ rise to 1.04\% at $\delta=10^\circ$, and (2) for $\tyc=1/2$ the error
in $\sigma_t$ is $>$50\% at $\delta=10^\circ$. These cases correspond to the
optima for the imaging and time resolution, where the low-order terms are
minimised, and for which higher-order terms become more important, and it is
therefore not surprising that they do not perform well. The Taylor
expansion for $\sigma_t$ does not converge quickly and a large number of
additional terms in $\delta$ and $\frac{1}{4\tyf^2}$ would be required to
achieve the same accuracy as the other expansions presented here. If greater
accuracy is required for any of the expressions discussed here a Python notebook
that numerically integrates Equation~\ref{EQ::FUNCTIONAL} is included in the
online material\footnote{A script that can be adapted to produce extra terms in
the Taylor expansions using the MATLAB Symbolic Math Toolbox is also included.}.
The notebook also contains Python functions that implement
Equations~\ref{EQ::CENTROID_X}--\ref{EQ::VAR_T}.

\section{Discussion}
\label{SEC::DISCUSSION}

An analysis of the general case of spherical optical system with ``identical''
regularly-spaced infinitesimal mirror facets has been presented. The expressions
differ that those presented before in two ways: (1) generalizing the classical DC
constraint that the curvature radius and focal length are equal, and (2) accounting
for light loss around the edges of the facets whose normals are not in general
aligned with the rays. In addition to presenting expressions for the moments of
the image on the focal plane, expansions for the effective area,
which accounts for light losses, and the dispersion of the distribution of
propagation times through the system have been presented.

In the case of a pure DC design these new expansions differ to some degree from 
those of \citetalias{REF::VASSILIEV::APP2007} (as corrected by
\citetalias{REF::BRETZ_RIBORDY::DC2013}), due to the effect of the rays lost at
the facet edges. For example, with $\tyc=1$ Equation~\ref{EQ::VAR_X} can be
approximated as,
$ \frac{\langle\Delta x^2\rangle}{F^2} =
\frac{\delta^2}{1024\tyf^4}\left(1+\frac{0.3073}{\tyf^2}\right) +
\frac{\delta^4}{16\tyf^2}\left(1+\frac{0.3264}{\tyf^2}\right)$,
while the equivalent expression from \citetalias{REF::VASSILIEV::APP2007}
is,
$ \frac{\langle\Delta x^2\rangle}{F^2} =
\frac{\delta^2}{1024\tyf^4}\left(1+\frac{0.3125}{\tyf^2}\right) +
\frac{\delta^4}{16\tyf^2}\left(1+\frac{0.3646}{\tyf^2}\right)$.
For the parabolic case, Equation~\ref{EQ::VAR_X} (with $\tyc=1/2$) agrees with 
that of \citetalias{REF::SCHLIESSER::THIRD_ORDER::APP2005} to the leading orders 
that they provide, both equalling
$ \frac{\langle\Delta x^2\rangle}{F^2} =
\frac{\delta^2}{512\tyf^2} + \frac{\delta^4}{16\tyf^2}$.

Writing the focal length as $F$ and the radius of curvature of the mirror sphere
as $R$, two privileged configurations are identified: $R=2/3F$ for which the
tangential coma term is minimised and equals the sagittal coma; and $R=2F$ which
minimises (largely eliminates) time dispersion, this latter giving a close
approximation to the parabolic surface.

The ratio $\tyc=F/R$ can therefore be used to trade off-axis imaging resolution
against time dispersion, by varying between the two optima of $\tyc=1/2$ and
$\tyc=3/2$. Using simulations of tessellated mirror systems
\citetalias{REF::SCHLIESSER::THIRD_ORDER::APP2005} discuss the same trade-off
and suggest an elliptical system with $R=0.85F$ as one of
interest\footnote{However, they use the classic DC alignment point at $z_A=2F$
rather than the value given by Equation~\ref{EQ::ALIGNMENT_PT}, for which
spherical aberration is suppressed, so their results (Figure~9) cannot be
approximated by Equation~\ref{EQ::VAR_X}.}. For the CTA MST a system with
$R=1.2F$ was chosen to improve the time resolution at the cost of slightly
degraded imaging resolution \citep{REF::GARCZARCZYK::MST::ICRC2015}.

The cost of a DC-like telescope is largely determined by the size of the
aperture, $D$, which determines the total mirror surface area, and the focal
length, which determines the size of the camera and the cost of the structure
required to support it. The cost associated with changing the curvature radius
from the DC case of $R=F$ comes from (1) accommodating a larger maximum sag (if
$R<F$), and (2) having mirror facets with different optimal focal distances. The
maximal sag for a telescope of diameter $D$ is approximately
$z_{max}\approx\frac{D^2}{8R}$. For example, for a system with $D=10$\,m and
$F=12$\,m this amounts to $z_{max}=1.09$\,m for the DC case ($R=12$\,m),
$z_{max}=0.53$\,m for the parabolic case ($R=24$\,m), and $z_{max}=1.76$\,m for
the minimum coma configuration ($R=8$\,m)\footnote{A mirror support structure
with this curvature would resemble that of the Whipple 10\,m, the prototypical
IACT, which had $D=10$\,m and $R=7.3$\,m
\citep{REF::FAZIO::WHIPPLE_10M::CJP1968}.}. The difference in cost of the
structures with this range of maximum sag, $(0.53,1.76)$\,m, is likely small
compared to the total cost of the system; this parameter can be used
to tune the performance of the system without impacting the cost excessively.

This note does not study the effects of finite mirror facet size. The size
of these effects depends primarily on the facet $f$-number, which is usually
quite large, the surface quality of the facets, the distribution of focal
lengths across the support structure, and deformations of the structure as it
moves. The shape and layout of the facets on the telescope play secondary roles.
The phase space for these effects is therefore large and there is no simple
equation to describe the combined PSF from a faceted telescope, although
\citetalias{REF::BRETZ_RIBORDY::DC2013} provide reasonable fits to simulations
in the DC case. Ray-tracing simulations are generally required to calculate the
detailed response of the system.


\bibliographystyle{elsarticle-harv}
\bibliography{shortreferences}
\label{SEC::REFERENCES}
%
% %% else use the following coding to input the bibitems directly in the
% %% TeX file.
% \end{thebibliography}
\end{document}